\nonstopmode
\documentstyle{article}

\def\underline#1{{\bf #1}}

\begin{document}

\def\beqn{\begin{eqnarray}{}}
\def\eeqn{\end{eqnarray}}
\def\oas#1{\hbox{${\cal O}(\alpha_s^{#1})$}}
\def\gsim{\lower0.5ex\hbox{$\stackrel{>}{\sim}$}}
\def\lsim{\lower0.5ex\hbox{$\stackrel{<}{\sim}$}}
\def\ff#1{\ifmmode{{}^{#1}F}\else{${}^{#1}F$}\fi}
\def\kk#1{\ifmmode{{}^{#1}{\rm K}}\else{${}^{#1}{\rm K}$}\fi}
\def\alphas{\alpha_s}
\def\eq#1{Equation~(\ref{eq:#1})}
\def\fig#1{Figure~\ref{fig:#1}}
\def\mq{M_Q}
\def\MSbar{\hbox{$\overline{\rm MS}$}}

\showboxdepth=1

\input epsf.def
\epsfverbosetrue

\newtoks\test
\test{800}

\def\vfigspace{\vspace{10pt}}

\def\figbint{
\begin{figure}[t]
\vfigspace
 \epsfscale=\the\test
  \epsfbox{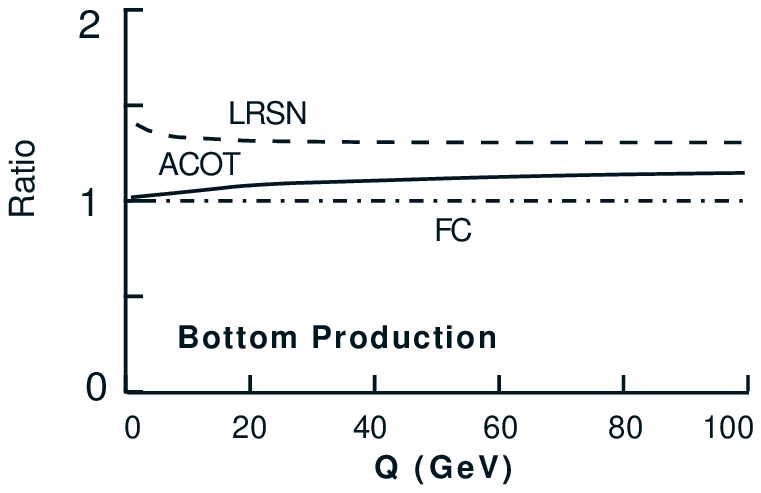}
      \caption{
The structure function $F_2(Q^2)$ scaled by the flavor creation (FC) $F_2(Q^2)$ 
for  $b$-quark production.
$F_2(x,Q^2)$ has been integrated over $x$ from $x=10^{-4}$ to $x_{max}$ as defined
in \protect\eq{G}. 
   }
   \label{fig:bint}
\end{figure}
}
\def\figbmui{
\begin{figure}[t]
\vfigspace
 \epsfscale=\the\test
  \epsfbox{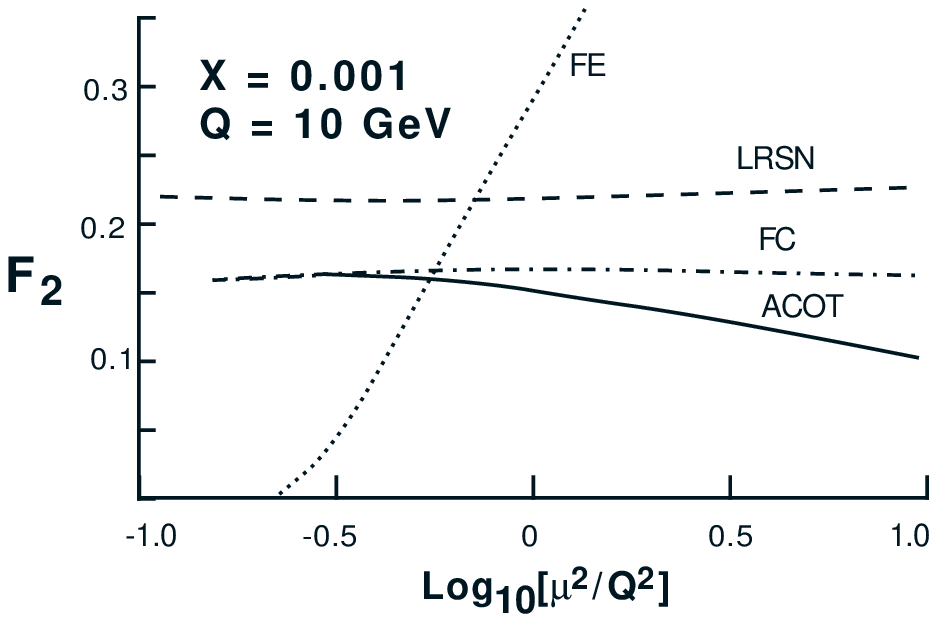}
      \caption{
The $\mu$-dependence of $F_2(x,Q^2)$ with $\{x,Q\}=\{10^{-3},10 \, GeV\}$
for $b$-quark production.
   }
   \label{fig:bmui}
\end{figure}
}
\def\figbmuii{
\begin{figure}[t]
\vfigspace
 \epsfscale=\the\test
  \epsfbox{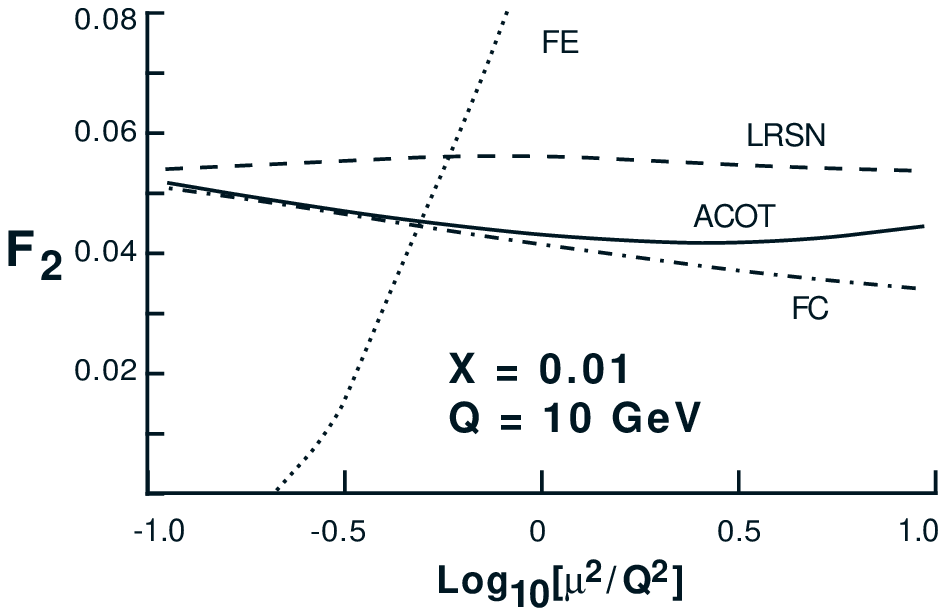}
      \caption{
The $\mu$-dependence of $F_2(x,Q^2)$ with $\{x,Q\}=\{10^{-2},10 \, GeV\}$
for $b$-quark production.
   }
   \label{fig:bmuii}
\end{figure}
}
\def\figbmuiii{
\begin{figure}[t]
\vfigspace
 \epsfscale=\the\test
  \epsfbox{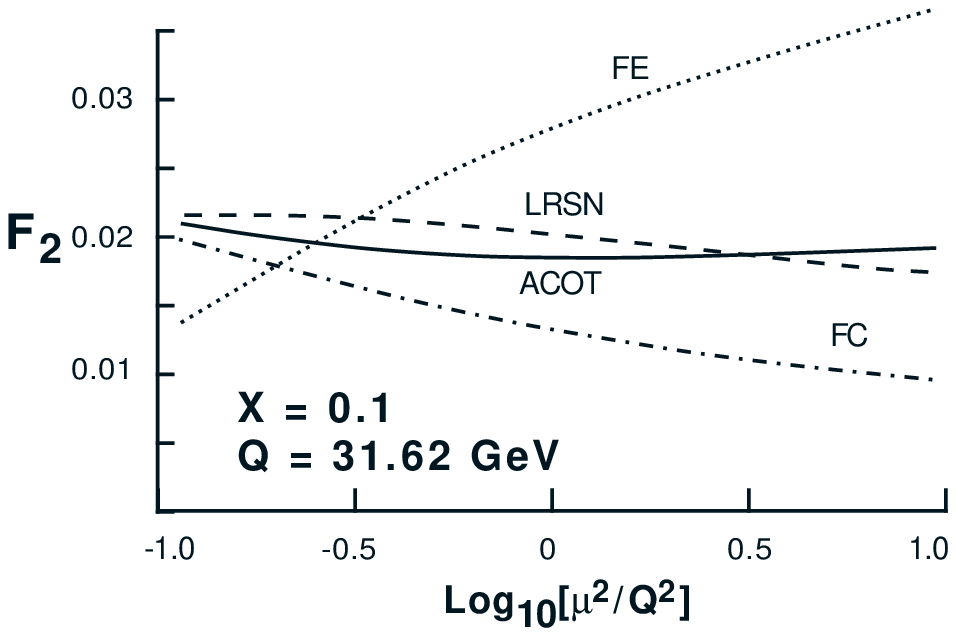}
      \caption{
The $\mu$-dependence of $F_2(x,Q^2)$ with $\{x,Q\}=\{10^{-1},31.62 \, GeV\}$
for $b$-quark production.
   }
   \label{fig:bmuiii}
\end{figure}
}
\def\figbxi{
\begin{figure}[t]
\vfigspace
 \epsfscale=\the\test
  \epsfbox{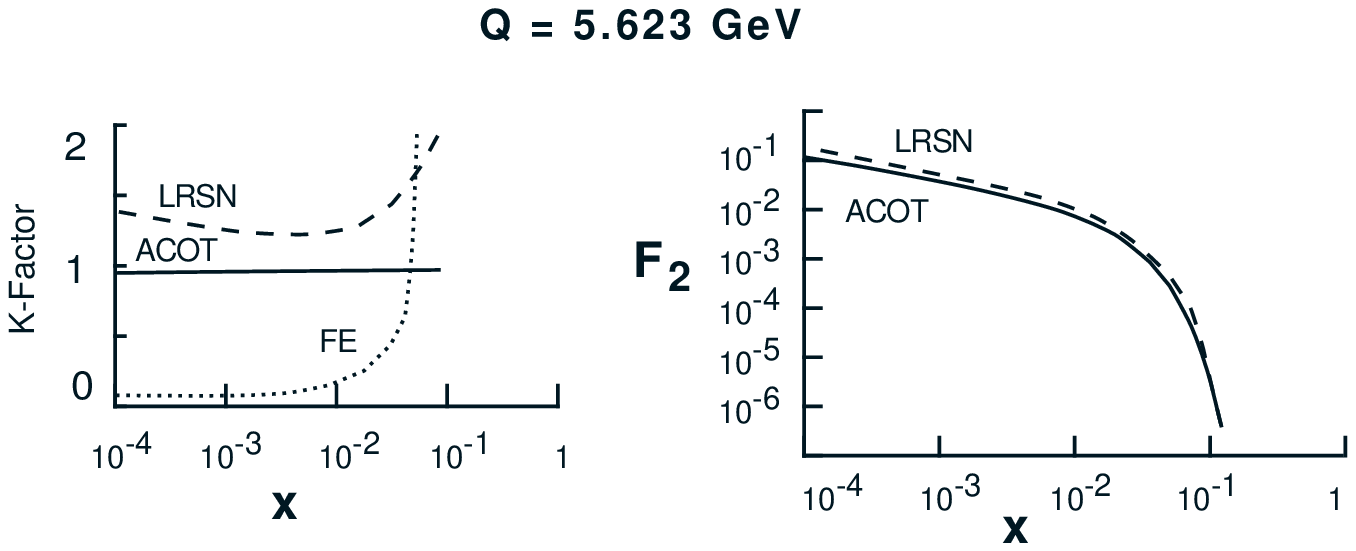}
      \caption{
\protect\ref{fig:bxi}a) the K-factor, and 
\protect\ref{fig:bxi}b) the structure function $F_2(x,Q^2)$
for $b$-production at $Q= 5.623 GeV$.
   }
   \label{fig:bxi}
\end{figure}
}
\def\figbxii{
\begin{figure}[t]
\vfigspace
 \epsfscale=\the\test
  \epsfbox{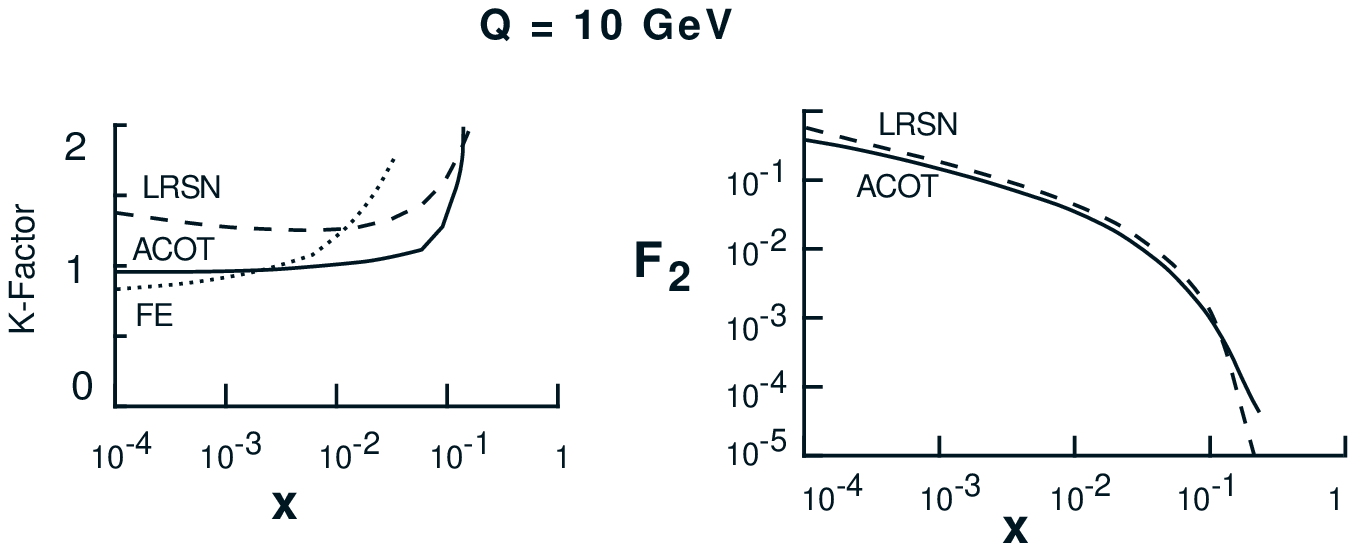}
      \caption{
\protect\ref{fig:bxii}a) the K-factor, and 
\protect\ref{fig:bxii}b) the structure function $F_2(x,Q^2)$
for $b$-production at $Q= 10 GeV$.
   }
   \label{fig:bxii}
\end{figure}
}
\def\figbxiii{
\begin{figure}[t]
\vfigspace
 \epsfscale=\the\test
  \epsfbox{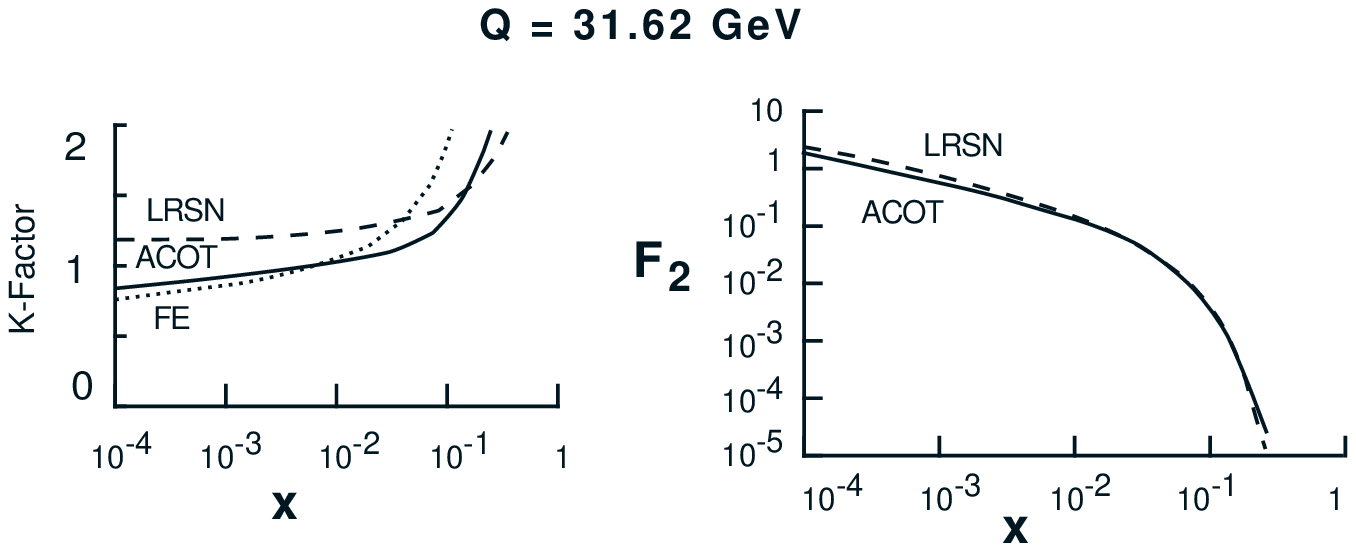}
      \caption{
\protect\ref{fig:bxiii}a) the K-factor, and 
\protect\ref{fig:bxiii}b) the structure function $F_2(x,Q^2)$
for $b$-production at $Q= 31.62 GeV$.
   }
   \label{fig:bxiii}
\end{figure}
}
\def\figcint{
\begin{figure}[t]
\vfigspace
 \epsfscale=\the\test
  \epsfbox{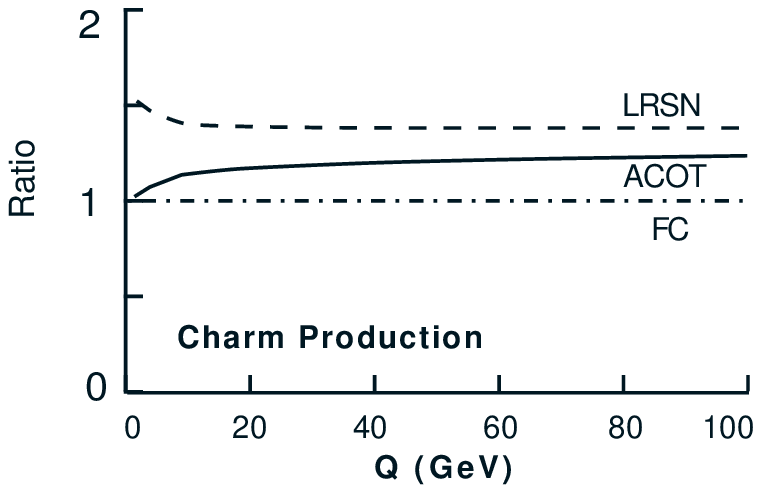}
      \caption{
The structure function $F_2(Q^2)$ scaled by the flavor creation (FC) $F_2(Q^2)$ 
for  $c$-quark production.
$F_2(x,Q^2)$ has been integrated over $x$ from $x=10^{-4}$ to $x_{max}$ as defined
in \protect\eq{G}. 
   }
   \label{fig:cint}
\end{figure}
}
\def\figcmui{
\begin{figure}[t]
\vfigspace
 \epsfscale=\the\test
  \epsfbox{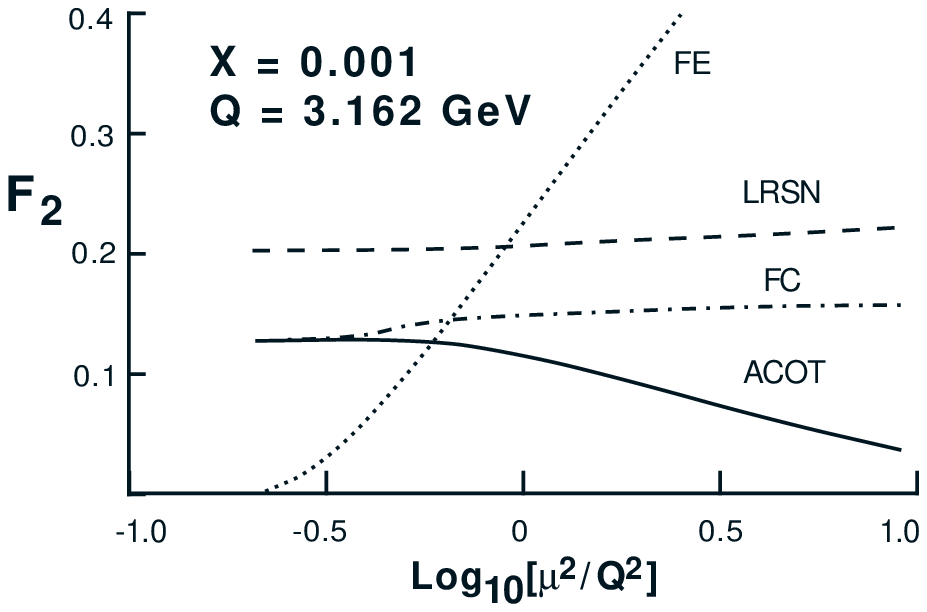}
      \caption{
The $\mu$-dependence of $F_2(x,Q^2)$ with $\{x,Q\}=\{10^{-3}, 3.162 \, GeV\}$
for $c$-quark production.
   }
   \label{fig:cmui}
\end{figure}
}
\def\figcmuii{
\begin{figure}[t]
\vfigspace
 \epsfscale=\the\test
  \epsfbox{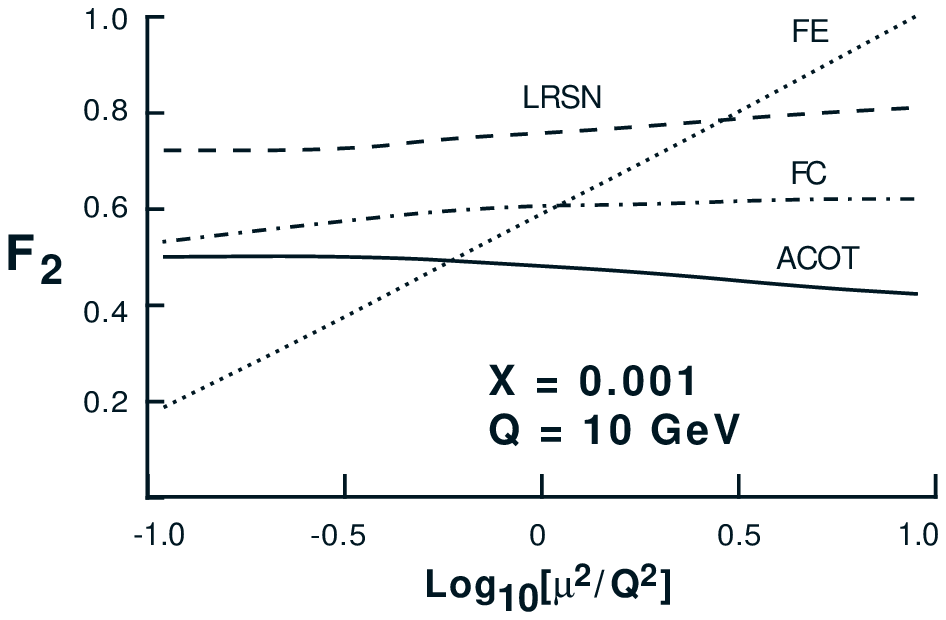}
      \caption{
The $\mu$-dependence of $F_2(x,Q^2)$ with $\{x,Q\}=\{10^{-3}, 10 \, GeV\}$
for $c$-quark production.
   }
   \label{fig:cmuii}
\end{figure}
}
\def\figcmuiii{
\begin{figure}[t]
\vfigspace
 \epsfscale=\the\test
  \epsfbox{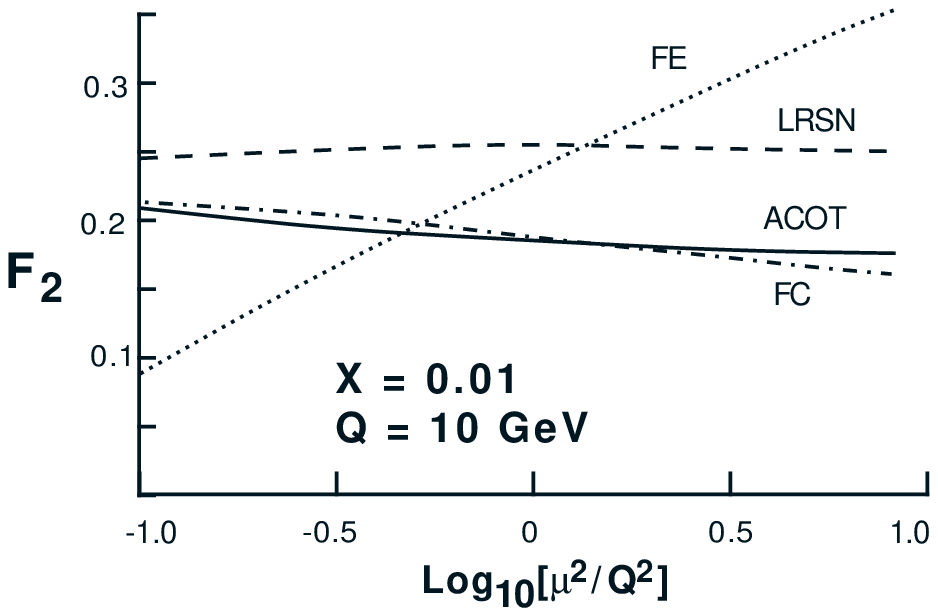}
      \caption{
The $\mu$-dependence of $F_2(x,Q^2)$ with $\{x,Q\}=\{10^{-2}, 10 \, GeV\}$
for $c$-quark production.
   }
   \label{fig:cmuiii}
\end{figure}
}
\def\figcmuiv{
\begin{figure}[t]
\vfigspace
 \epsfscale=\the\test
  \epsfbox{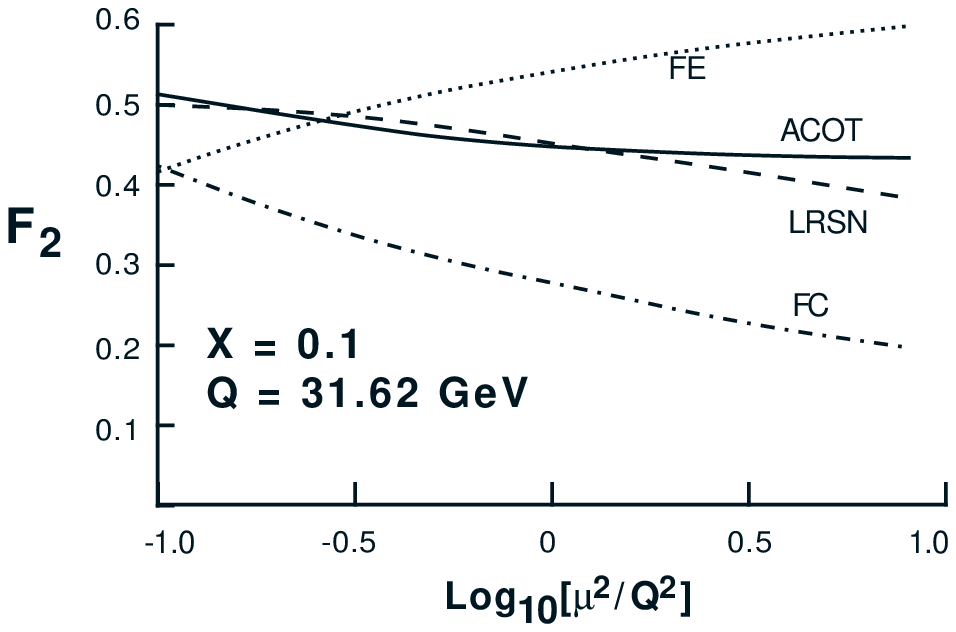}
      \caption{
The $\mu$-dependence of $F_2(x,Q^2)$ with $\{x,Q\}=\{10^{-1}, 31.62 \, GeV\}$
for $c$-quark production.
   }
   \label{fig:cmuiv}
\end{figure}
}
\def\figcxi{
\begin{figure}[t]
\vfigspace
 \epsfscale=\the\test
  \epsfbox{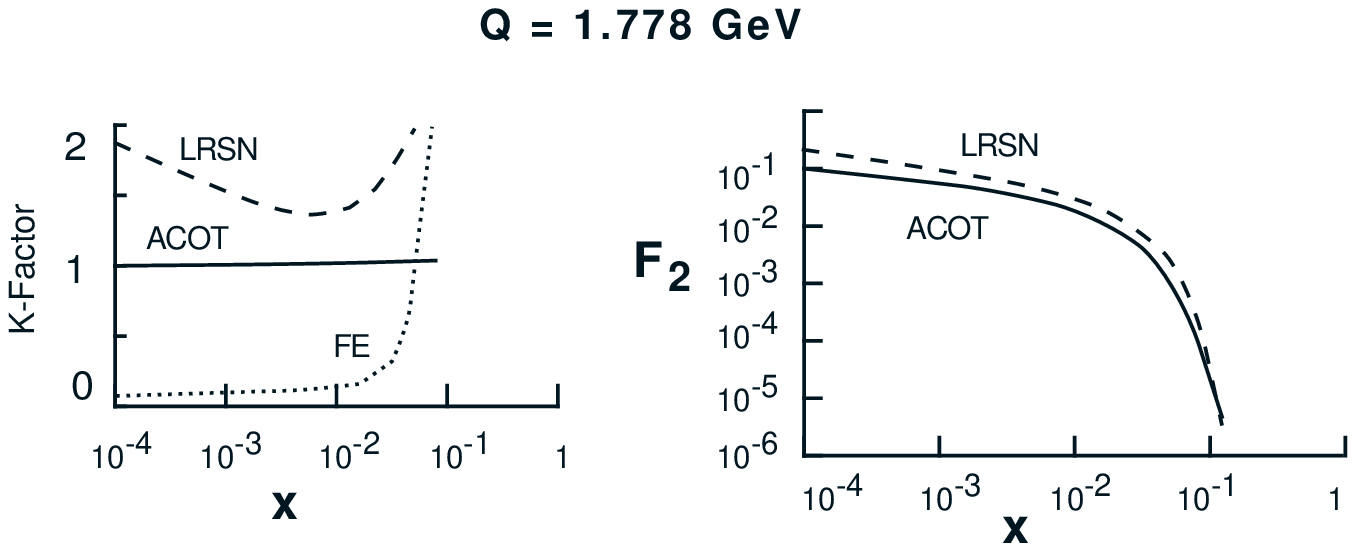}
      \caption{
\protect\ref{fig:cxi}a) the K-factor, and 
\protect\ref{fig:cxi}b) the structure function $F_2(x,Q^2)$
for $c$-production at $Q= 1.778 GeV$.
   }
   \label{fig:cxi}
\end{figure}
}
\def\figcxii{
\begin{figure}[t]
\vfigspace
 \epsfscale=\the\test
  \epsfbox{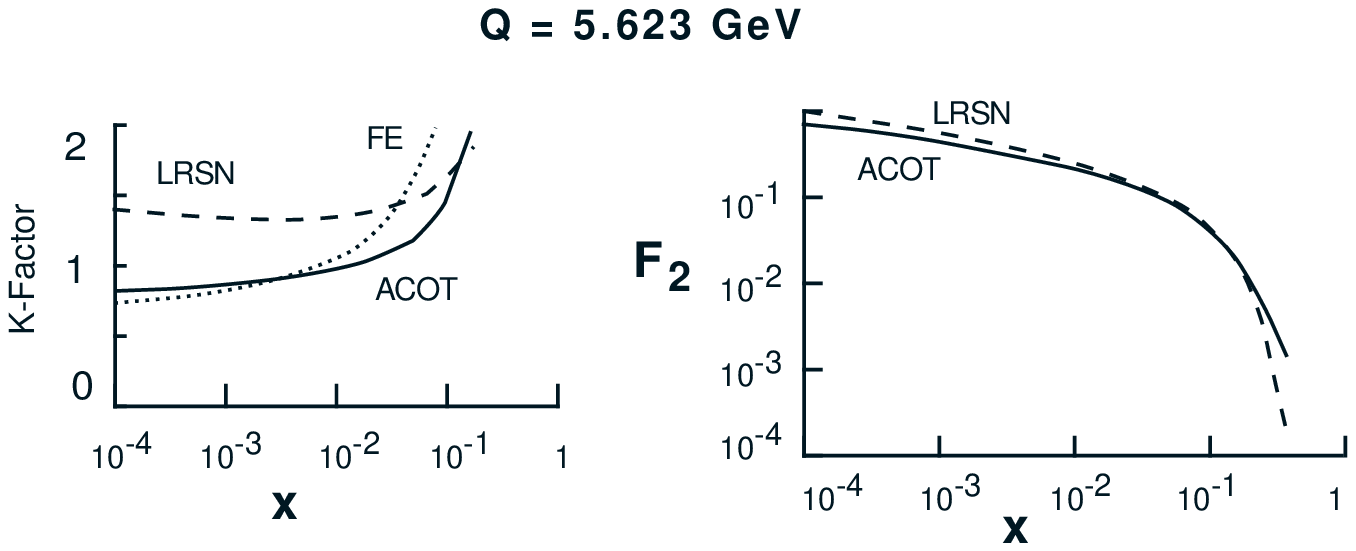}
      \caption{
\protect\ref{fig:cxii}a) the K-factor, and 
\protect\ref{fig:cxii}b) the structure function $F_2(x,Q^2)$
for $c$-production at $Q= 5.623 GeV$.
   }
   \label{fig:cxii}
\end{figure}
}
\def\figcxiii{
\begin{figure}[t]
\vfigspace
 \epsfscale=\the\test
  \epsfbox{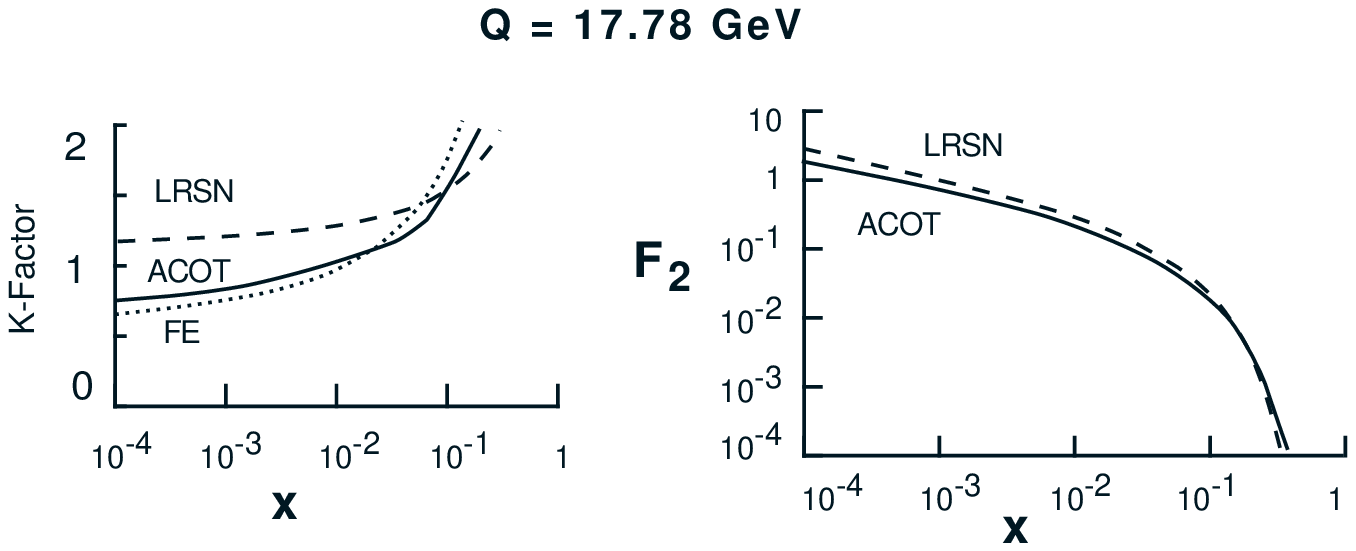}
      \caption{
\protect\ref{fig:cxiii}a) the K-factor, and 
\protect\ref{fig:cxiii}b) the structure function $F_2(x,Q^2)$
for $c$-production at $Q= 17.78 GeV$.
   }
   \label{fig:cxiii}
\end{figure}
}
\def\figxfigi{
\begin{figure}[t]
\vfigspace
 \epsfscale=\the\test
  \epsfbox{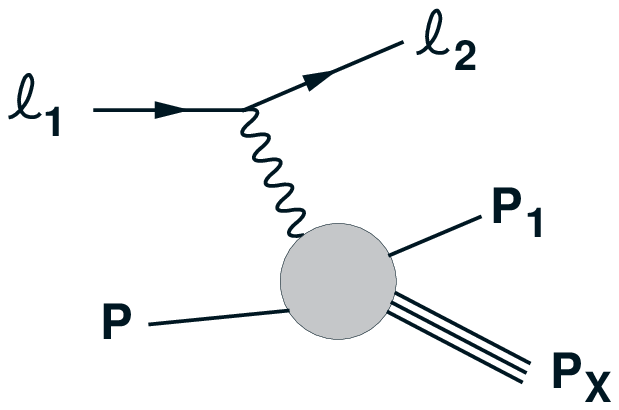}
      \caption{
The basic semi-inclusive deeply inelastic scattering process,
\hbox{
$\ell_1(\ell_1) + N(P) \longrightarrow \ell_2(\ell_2) + Q(p_1) + X(P_X)$.}
   }
   \label{fig:xfigi}
\end{figure}
}
\def\figxfigii{
\begin{figure}[t]
\vfigspace
 \epsfscale=\the\test
  \epsfbox{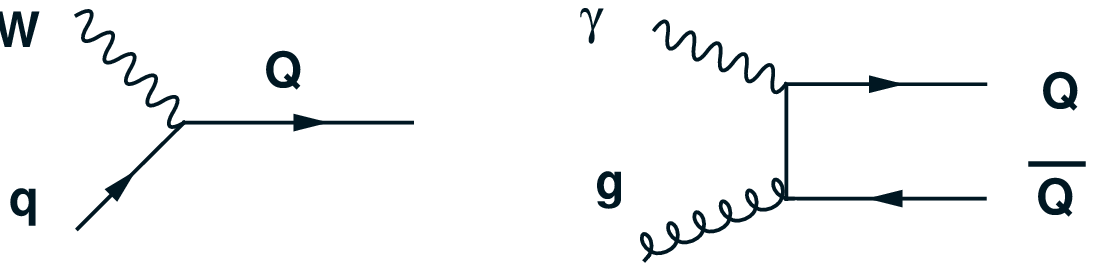}
      \caption{
\protect\ref{fig:xfigii}a)
The \oas0 flavor excitation (FE) process, shown here for the charged-current case.
\protect\ref{fig:xfigii}b)
The \oas1 flavor creation   (FC) process, shown here for the neutral-current case.
   }
   \label{fig:xfigii}
\end{figure}
}
\def\figxfigiv{
\begin{figure}[t]
\vfigspace
 \epsfscale=\the\test
  \epsfbox{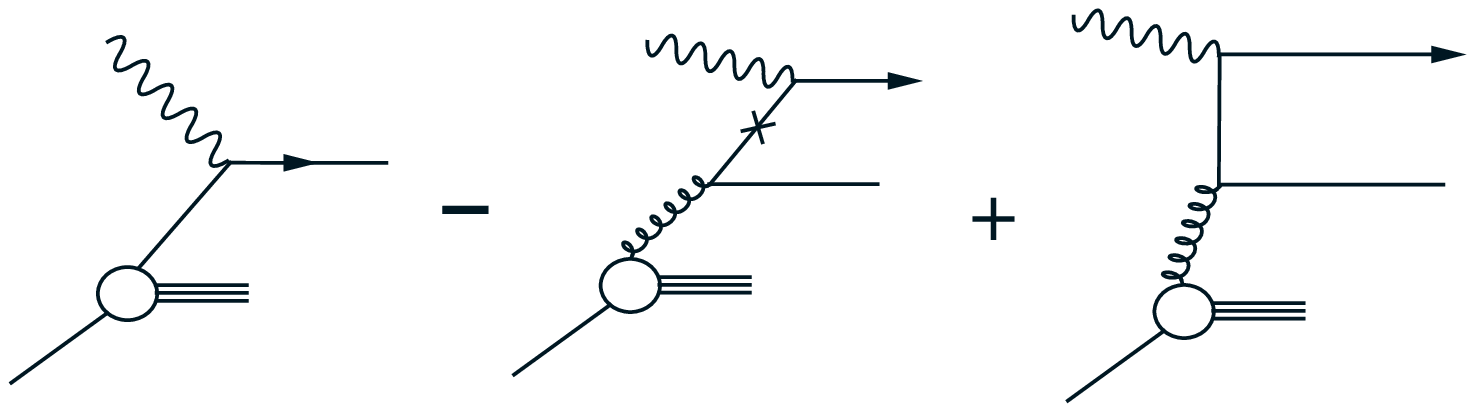}
      \caption{
Pictorial representation of the variable flavor scheme (VFS) calculation 
of ACOT.
\protect\ref{fig:xfigiv}a) The \oas0 flavor excitation (FE) contribution.
\protect\ref{fig:xfigiv}b) The schematic representation of the subtraction term. 
The $\times$ on the intermediate quark line indicates a convolution between the 
\oas1 $g\to q$ splitting (lower portion of the diagram) with the \oas0 process 
(upper portion of the diagram).
\protect\ref{fig:xfigiv}c) The \oas1 flavor creation   (FC) contribution.
   }
   \label{fig:xfigiv}
\end{figure}
}
\def\figxfigiii{
\begin{figure}[t]
\vfigspace
 \epsfscale=\the\test
  \epsfbox{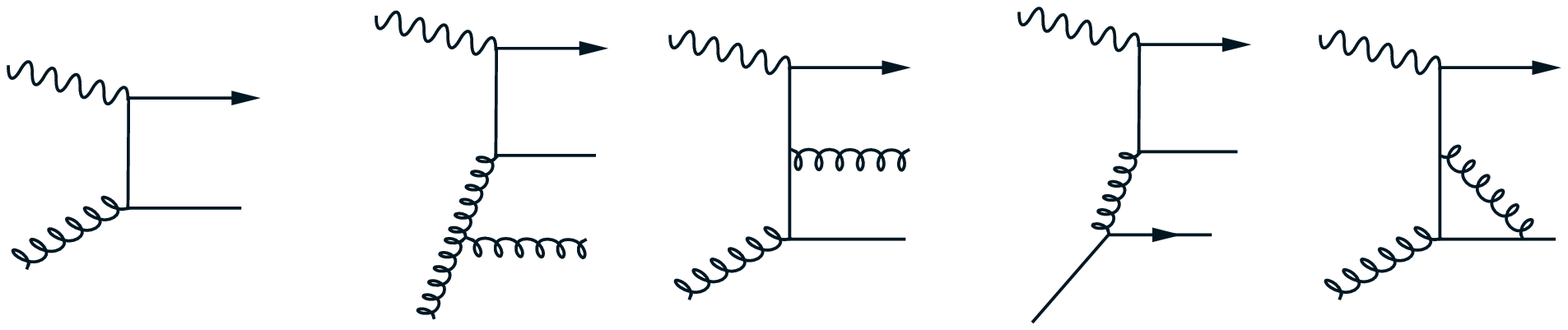}
      \caption{
Pictorial representation of the fixed flavor scheme (FFS) calculation of LRSN.
\protect\ref{fig:xfigiii}a) The \oas1 flavor creation   (FC) contribution.
\protect\ref{fig:xfigiii}b), 
\protect\ref{fig:xfigiii}c), 
\protect\ref{fig:xfigiii}d), and
\protect\ref{fig:xfigiii}e) An illustrative subset of the complete set
of \oas2 diagrams contributing to the complete LRSN calculation.
   }
   \label{fig:xfigiii}
\end{figure}
}
\def\figxfigv{
\begin{figure}[t]
\vfigspace
 \epsfscale=\the\test
  \epsfbox{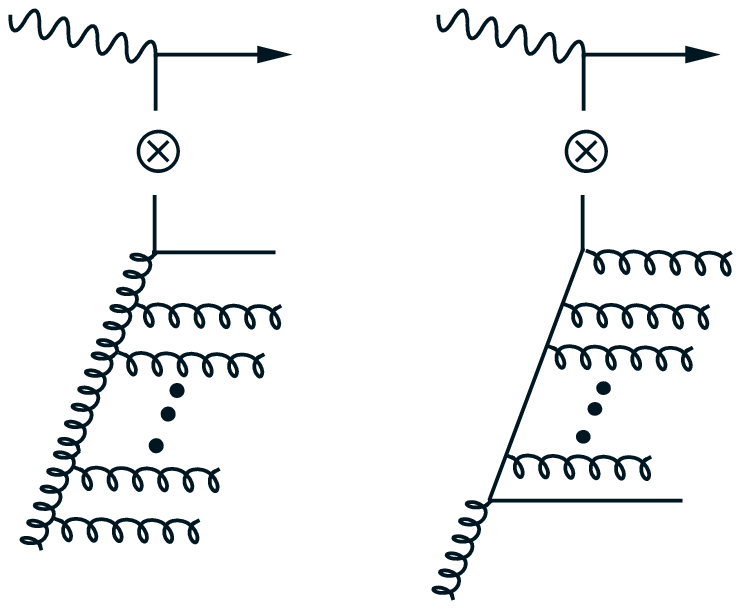}
      \caption{
 An illustrative subset of the class of diagrams which are 
summed by the QCD evolution of the heavy quark PDF in the variable flavor
scheme (VFS) calculation of ACOT. 
The $\otimes$ indicates a convolution between the heavy quark PDF
 (lower portion of the diagram) and the \oas0 process 
(upper portion of the diagram).
   }
   \label{fig:xfigv}
\end{figure}
}


\null
\vfil

\begin{center}
\begin{tabular}{l}
August 1994 \\
\end{tabular}
    \hfill
\begin{tabular}{l}
SMU-HEP/94-21 \\
CTEQ 94-07 \\
hep-ph/9409208
\end{tabular}
\\[1cm]

{\bf{\LARGE  Leptoproduction of Heavy Quarks \\
in the  Fixed and Variable Flavor Schemes}}
\\[.5in]

 {\large
    Fredrick I. Olness\footnote{SSC Fellow}
   and Stephan T. Riemersma
  }
 \\[0.5in]
  Southern Methodist University,
   Dallas, Texas 75275
\end{center}
\vfil

\begin{abstract}

We compare the results of the fixed-flavor scheme calculation of
Laenen, Riemersma,  Smith and van Neerven with the
variable-flavor scheme calculation of Aivazis, Collins, Olness
and Tung for the case of neutral-current (photon-mediated)
heavy-flavor (charm and bottom) production.  Specifically, we
examine the features of both calculations  throughout phase space
and compare the structure function $F_2(x,Q^2)$.   We also
analyze the dependence of $F_2$ on the mass  factorization scale
$\mu$. We find that the former is most applicable near threshold,
while the latter works well for asymptotic $Q^2$.  The validity
of each  calculation in the intermediate region is dependent upon
the $x$ and $Q^2$ values chosen.

\end{abstract}

 \vfil

\newpage
\section{Introduction}

Several experimental groups \cite{dis}
have studied the semi-inclusive deeply
inelastic scattering (DIS) process for heavy-quark production (\fig{xfigi})
\beqn
\ell_1(\ell_1) + N(P) \rightarrow \ell_2(\ell_2) + Q(p_1) + X(P_X).
\label{eq:dis}
\eeqn
Most analyses of this process assume that the hadron is comprised only of
the massless  gluon ($g$), up ($u$), down ($d$)  and strange ($s$) quarks,
while the charm ($c$), bottom ($b$), and top ($t$) quarks are treated as
massive objects which are strictly external to the hadron.
\figxfigi

This view of the heavy quarks as external to the hadron
is appropriate when the energy scale of the process $\mu_{\rm phy}$
(for example the center-of-mass energy $\sqrt{s}$)
is not large compared to the mass of the heavy quark,
{\it i.e.} $M_Q \lsim \mu_{\rm phy}$.
For most fixed-target facilities, this condition holds for the $c$, $b$
and $t$ quarks~\cite{sobt}.
We are therefore justified in excluding $c$, $b$, and $t$ as
constituents of the hadron in the QCD-improved quark-parton model (QPM)
for this case.

With new data from HERA, the electron-proton collider at DESY,
we can investigate the DIS process in a very different kinematic range from
that available at fixed-target experiments \cite{sobt}.  In this
new realm, the important question is:
Should the $c$ and $b$ quarks be considered as partons, or as heavy
objects extrinsic to the hadron?
Given that HERA extends the kinematic reach of the DIS process by
two orders of magnitude,
we can not expect our assumptions that were valid for fixed-target processes
to hold in a completely different kinematic regime.

Aivazis, Collins, Olness and Tung (ACOT) have discussed this issue at length
in reference~\cite{acot} and approach the problem by invoking
the {\it variable flavor scheme} (VFS), which varies the number of partons
according to the relevant energy scale $\mu_{\rm phy}$.
 The fundamental physical insight to the VFS is that in the region
$M_Q \,  \gg \, \mu_{\rm phy}$, the heavy quark should be {\it excluded}
as a constituent of the
hadron as it is kinematically inaccessible.
However, when
$M_Q \, \ll \, \mu_{\rm phy}$ the heavy quark should be {\it included} as a
parton since $\mq$ is insignificant compared to $\mu_{\rm phy}$.
Although the physics is unambiguous in these kinematic extremes,
most experimental data lies in between these clear-cut regions.
In the intermediate region, the renormalization scheme of Collins,
Wilczek and Zee (CWZ) \cite{cwz}
provides a well-defined transition between these two
extreme kinematic domains.

The issue of what constitutes a parton also points to an
inconsistency  between traditional charged-current and neutral-current
heavy-quark production calculations~\cite{acot}.
When considering charged-current processes, one begins with  the purely
electroweak process $W + q \to Q$, as shown in \fig{xfigii}a.
For neutral-current processes, the traditional approach is to begin
with  the \oas1 process $\gamma^* + g \to Q + \overline{Q}$,  as shown in
\fig{xfigii}b.\footnote{Due to the traditional inconsistency between the
neutral and  charged processes discussed above, the terminology leading-order
and next-to-leading-order is ambiguous. Therefore, we denote the subprocesses
according to the power of $\alpha_s$.}
As we work in the new kinematic regime spanned by HERA, the concept of
a `heavy' quark becomes a relative term depending upon the magnitudes of
the kinematic variables involved.
 Traditional distinctions between the charged-current and neutral-current
calculations should vanish as the characteristic energy scale becomes
significantly larger than the heavy-quark mass.  ACOT implements the CWZ
renormalization scheme and treats both charged-current and neutral-current
heavy-flavor production in the consistent fashion of beginning both
calculations
at \oas0.
\figxfigii

Laenen, Riemersma, Smith and van Neerven (LRSN) have calculated heavy-quark
production for DIS photon exchange, beginning with the \oas1 photon-gluon
fusion process and including the
complete \oas2 radiative corrections in reference~\cite{lrsn}.  \fig{xfigiii}
displays some of the relevant ${\cal O}(g^2)$ tree-level and ${\cal O}(g^3)$
virtual Feynman diagrams involved, where $g$ is the QCD coupling strength
and is given by
$\alpha_s = g^2/4 \pi$.  LRSN calculates from the viewpoint that the
produced heavy quark is generated from an initial-state gluon or lighter
quark.  For example, in producing $c$ quarks, LRSN assumes the parton
initiating the heavy-quark production process to be the $g$, $u$, $d$ or $s$
quark.  The same tenet holds for $b$ production, only the $c$ quark would
be included as a massless initial-state parton as well.
\figxfigiii

Another feature of the LRSN calculation is that it provides additional
information on inclusive differential cross section distributions.  At \oas1,
the heavy quark and antiquark are produced back to back in the
$\gamma^*$-parton
center of momentum frame,  and at \oas2, the additional influences of the gluon
radiation and the $\gamma^* q$ channel have also been calculated in
\cite{lrsn2}.  LRSN gives additional insight into the differential
distributions, which is particularly useful from the experimental point of
view.

The subject of this paper is a vigorous comparison of the advantages
of each calculation, as well as a glimpse of future prospects including
a merging of the two calculations  to produce a three-order result which
should have excellent predictive power for heavy-flavor structure functions
at HERA.

\section{Kinematics}

While \fig{xfigi} shows the general deeply inelastic scattering (DIS) process,
we
 shall focus specifically on  neutral-current heavy-flavor production via a
photon
exchange\footnote{
The ACOT calculation with general masses and couplings applies to
both charged and neutral-current processes.}
 as described by the sub-process
$\gamma^*(q) + N(P) \longrightarrow Q(p_Q) + X(P_X)$.
We define
$Q^2 = -q^2$, where $q$ is the four momentum of the virtual photon exchanged,
$y= (P\cdot q) /( P \cdot \ell_1)$ is the fractional energy transfer,
and
$x = Q^2/(2P \cdot q)$ is the Bjorken $x$ variable.
 The cross section is obtained from the structure functions via
\beqn
\frac{d^2 \sigma}{dx \, dy} &=& \frac{2ME_1}{\pi}\frac{G_1G_2}{n_\ell}
\left\{g^2_{+l}
\left[ x F_1 \, y^2 + F_2 \, \left[(1 - y) -
\left(\frac{Mxy}{2E_1}\right)\right]
\right] \pm
g^2_{-l}
\left[x F_3 \, y \left(1 - \frac{y}{2}\right)\right]
\right\} \quad .
\label{acotsigma}
\eeqn
 $E_1$ is the energy of the incoming lepton,
 $M$ is the mass of the hadron being probed,
 $G_i = g^2_{B_i}/(Q^2 + M^2_{B_i})$ is a shorthand for the boson coupling and
the propagator,
  $n_\ell$ is the number of polarization states of the incoming lepton,
and
  $g^2_{\pm l} = g^2_{Ll} \pm g^2_{Rl}$, where $g_{Ll}$ and
 $g_{Rl}$ refer to the chiral couplings of the vector boson to the
leptons.\footnote{In this definition of the structure functions,
we have extracted the quark coupling ($g^2_{\pm l}$) and the
average over the incoming lepton polarization ($n_\ell$) so that
the same formula applies to both charged and neutral processes.}

\section{ACOT Calculation}

The ACOT calculation makes use of the \oas0 and \oas1 processes to obtain a
two-order result, {\it cf.} \fig{xfigii} and  \fig{xfigiv}. The \oas0
result uses the standard QCD evolution to resum the iterative
gluon and quark splittings to give the parton distribution function (PDF) of
the heavy quark.
\figxfigiv

The \oas0 term is straightforward, and is represented by
\beqn
\ff0 = F(\oas0) &=&  f_{N}^{Q_i} \otimes
\widehat{F}^{(0)}(\gamma^* Q_i \rightarrow Q_i)
\quad ,
\eeqn
where $f_{N}^{Q_i}$ denotes the parton distribution function
for a nucleon $N$ producing a quark $Q_i$, and the
$\widehat{F}^{(0)}(\gamma^* Q_i \rightarrow Q_i)$ is the lowest-order
partonic structure function for the initial-state heavy quark $Q_i$ to absorb
a virtual photon $\gamma^*$ and enter the final state.
We introduce the shorthand $F({\cal O}(\alpha_s^n))= \ff{n}$.  The
$\otimes$ represents the convolution which is the integral over the
momentum fraction carried by the initial state parton $i$
\beqn
f^i_N \otimes \widehat{F}  &=&
\int_{x}^{1} \
\frac{d\xi}{\xi} \
f^i_N \left(\frac{x}{\xi},\mu_f \right) \
\widehat{F}\left(\xi,\frac{Q}{\mu_f},\alpha_s(\mu_r)\right)
\quad .
\eeqn
where $\xi=k^+/P^+$ is the ratio of the ``$+$"-component of the parton
momentum $k^+$ to the hadron momentum $P^+$ in light cone coordinates,
 $\mu_r$ is the renormalization scale,
 and
 $\mu_f$ is the mass factorization scale.
For the remainder of this paper, we shall not distinguish
between $\mu_r$ and $\mu_f$, and we choose $\mu_r = \mu_f \equiv \mu$.
Although the partonic structure function $\widehat{F}$ and the evolution
of the PDF with mass factorization scale $\mu$ are calculable in perturbative
QCD,
the $\xi$-dependence of the PDF is not and must be
derived from experimental data.

Handling the collinear and infrared divergences of the $F(\oas1)$ requires
careful application of the factorization formula as illustrated in the ACOT
paper.  The complete \ff1 contribution in VFS is given by
\beqn
 \ff1_{\scriptscriptstyle V}
= F_{\scriptscriptstyle V}(\oas1)
= \ff{1u}_{\scriptscriptstyle V} - \ff{1s}_{\scriptscriptstyle V} &=&
f_{N}^{g} \otimes \widehat{F}^{(1)}(\gamma^* g \rightarrow Q_i \overline{Q}_i)
-
f_{N}^{g} \otimes f_{g}^{Q_i\,(1)}
\otimes \widehat{F}^{(0)}(\gamma^* Q_i \rightarrow Q_i)
\nonumber\\
\eeqn
$\ff{1u}_{\scriptscriptstyle V}$
is the `unsubtracted' \oas1 contribution, meaning the collinear
singularities are still present, and $\ff{1s}_{\scriptscriptstyle V}$
is the `subtraction' term
which cancels the singularities and renders \ff{1} finite.  The $V$ subscript
indicates the use of the VFS.
The reason for separating \ff{1} into these two pieces will
become clear in Section~5, where the K-factors are discussed.
Here, $f_{g}^{Q_i\,(1)}$ is the {\it perturbative} \oas1
splitting function for the process
$g \rightarrow Q_i$ and is given by
\beqn
f_{g}^{Q_i\,(1)}
&=&
{\alphas(\mu) \over 4\pi} \  C_R \
\ln\left( {\mu^2 \over \mq^2} \right) \
P_{g \rightarrow q} (x) \
\theta(\mu^2 - \mq^2),
\eeqn
where $P_{g \rightarrow q} (x) = t_F \, (1 -2x + 2x^2)$.
In SU(N), $C_R = (N^2 - 1)/2N$ and $t_F = 1/2$.
Note that $f_{g}^{Q_i\,(1)}=0$ when $\mu \leq \mq$, and evolves continuously
from zero for $\mu \geq \mq$.

\section{LRSN Calculation}

The LRSN calculation uses a {\it fixed flavor scheme} (FFS), which sets the
number of active flavors to a constant, regardless of the energy scale
$\mu_{\rm phy}$ of the
production process.  For example, when considering $b$ production the number
of light flavors would be four.

The lowest-order FFS \oas1 term is given by
\beqn
\ff1_{\scriptscriptstyle F} = F_{\scriptscriptstyle F}(\oas1)
&=&
f_{N}^{g} \otimes \widehat{F}^{(1)}(\gamma^* g \rightarrow Q_i \overline{Q}_i).
\eeqn
Mass factorization is not necessary at \oas1 in FFS since the mass of the
heavy quark is explicitly assumed to be non-zero throughout the calculation,
and the initial-state heavy quark contribution is not included.
Note the $\ff1_{\scriptscriptstyle F}$ contribution in this scheme
corresponds to the `unsubtracted' $\ff{1u}_{\scriptscriptstyle V}$
contribution of
VFS. We make use of this fact when we compare the K-factors in Section~8.

LRSN has computed the Feynman diagrams shown in \fig{xfigiii} to give the
complete
\oas2 result.  After cancelling the infrared divergences and performing the
necessary renormalization, the mass factorization is done according to

\beqn
\ff2 = F(\oas2) = \ff{2u} - \ff{2s}
&=&
f_{N}^{g} \otimes \widehat{F}^{(2)}(\gamma^* g \rightarrow Q_i \overline{Q}_i
g)
+
\sum_{a} f_{N}^{q_a} \otimes \widehat{F}^{(2)}(\gamma^* q_a\rightarrow Q_i
\overline{Q}_i q_a) - \, \ff{2s} \, ,
\nonumber\\
\eeqn
where all the contributions from initial-state partons light relative
to the produced heavy quark are summed.
As before, we recognize the first term as the `unsubtracted'
\ff{2u}(\oas2) contribution,
and the second term as the `subtraction' term \ff{2s}(\oas2) which removes the
collinear mass singularity from the radiation of a gluon by the incoming
massless quark or gluon.
Note the `subtraction' term does not subtract the $\mq \rightarrow 0$
singularity.  Mass factorization is not necessary as
$\mq^2/Q^2$ is explicitly  assumed to be non-zero and the initial-state heavy
quark contribution is assumed absent.  For additional details, see
references~\cite{lrsn} and~\cite{lrsn2}.
At \oas2 the FFS subscript is superfluous since no confusion is present at
this order.


\section{K-factors}

The two calculations each include a different subset of the complete set
of higher-order corrections.  The comparison of the two approaches will
determine to what extent these calculations pick up similar higher-order
contributions, and  can be used to estimate the
magnitude of the corrections not included by either approach.
To facilitate this comparison, we shall focus on the $F_2$ K-factors
of each calculation as a function of $x$ and $Q^2$.
For the LRSN calculation, we define the K-factor to be
\beqn
K_{LRSN}
&=&
\frac{\ff{1}_{\scriptscriptstyle F} + \ff2 }{\ff{1}_{\scriptscriptstyle F}}
\equiv
\frac{\ff{1u}_{\scriptscriptstyle V} + \ff2 }{\ff{1u}_{\scriptscriptstyle V}}
\, ,
\label{eq:kfacLRSN}
\eeqn
where the $K_{LRSN}$ indicates the K-factor is coming from the LRSN \oas2
contributions.

For ACOT, we define the K-factor as
\beqn
K_{ACOT}
&=& \frac{\ff0_{\scriptscriptstyle V}  + \ff1_{\scriptscriptstyle V} }{
\ff{1u}_{\scriptscriptstyle V}}
\equiv
\frac{ \ff{1u}_{\scriptscriptstyle V} + (\ff0_{\scriptscriptstyle V} -
\ff{1s}_{\scriptscriptstyle V})}{ \ff{1u}_{\scriptscriptstyle V}} \, ,
\label{eq:kfacACOT}
\eeqn
where $K_{ACOT}$ indicates the K-factor is coming from the ACOT \oas0
contribution.
$K_{ACOT}$ is defined in this manner because it is the
$\ff{1}_{\scriptscriptstyle F} \equiv \ff{1u}_{\scriptscriptstyle V}$
contribution which is common to both calculations. Therefore, we use this
common term to set the scale of comparison in the denominator of the K-factor.
Secondly, in the threshold region ($Q \simeq \mq$), the production cross
section is dominated by the \oas1 process. Viewing the \oas0 process as a
perturbation on the dominant \oas1 process, we see \eq{kfacACOT} is the
natural definition of the K-factor.

Both of these points become clear when the second form of \eq{kfacACOT} is
compared with \eq{kfacLRSN}.
The higher-order contributions for the LRSN calculation are contained in the
\ff2 term, whereas $\ff0 - \ff{1s}$ yields the contribution for the
ACOT calculation.

\section{Parton Distributions for Heavy Quarks: The VFS Approach}

The VFS calculation of ACOT uses the CWZ renormalization scheme to
incorporate the heavy quark into the QPM. The fundamental idea here is that at
low energy scales ($Q \ll \mq$), we do not want to treat the heavy quark as a
constituent of the hadron.  However, at high energy scales ($Q \gg \mq$), the
mass of the `heavy' quark is negligible and we should therefore treat this
quark on the same footing as the other massless partons. The concept that
there should be a democracy among the quarks in the limit
$Q/\mq \rightarrow \infty$
forces us to introduce the heavy quark as a parton at some intermediate energy
scale which is typically related to the quark mass. The choice of
this scale is intimately connected to the choice of renormalization
scheme \cite{ct}, \cite{sq}.

The motivation to include the heavy quark as a constituent of the hadron
is more than aesthetic.  When computing physical cross sections in the context
of perturbation theory, we find that our perturbation expansion is not simply
an expansion in powers of $\alphas$. As we compute higher-order subprocesses,
we gain logarithms involving the various energy scales in the problem (such as
$\ln(Q/\mq)$). Therefore, we find that our perturbation series is actually an
expansion in $\alphas \ln(Q/\mq)$. In the limit $Q/\mq\rightarrow \infty$, we
will clearly have a divergent series unless we can resum these
logarithmic terms.

The QCD evolution of the parton distribution functions does precisely this
resummation for the partons. It sums an infinite set of quark and gluon
splittings, thereby taking all such logarithmic contributions into account.
To be perfectly clear, these resummations are done in both LRSN and ACOT
for the light quarks and the gluon.  The only difference is in the way the
heavy quark is treated.  For ACOT, it is incorporated into the PDF, and for
LRSN, it only enters in the final state as a product of an interaction of
a light quark or gluon with the virtual photon.

The goal of the ACOT calculation is to use this same technique to resum the
multiple splittings arising from the QCD evolution of the heavy-quark
distribution.  The result is to reorganize the perturbation expansion such
that the singular $\alphas \ln(Q/\mq)$ terms are split into
an $\alphas \ln(Q/\mu)$ term which remains in the hard scattering Wilson
coefficient and
an $\alphas \ln(\mu/\mq)$ term which is absorbed into the heavy-quark
PDF.
The end result is that the perturbation series then becomes an expansion in
powers of $\alphas \ln(Q/\mu)$, and we retain the freedom to adjust $\mu$ to
optimize the convergence of the series.

This is clearly the correct approach in the asymptotic limit $Q/\mq \rightarrow
\infty$
but how well does it work in the kinematic regime of present
accelerators? The only way to answer this question is to compare ACOT with a
separate calculation such as LRSN.

\section{Kinematic Regimes}

\subsection{Collinear/On-Shell vs. Large $P_T$/Virtual}

The VFS approach  (ACOT) resums an infinite
subset of diagrams (\fig{xfigiv}) within the PDF of the heavy
quark, {\it cf.}, via the flavor excitation (FE) (\fig{xfigiii}a).
Because the heavy quark is treated as a parton for this set of contributions,
the phase space of the heavy quark is restricted to the on-shell collinear
region.  In the asymptotic limit $\mu \to \infty$, the  on-shell collinear
region is the dominant region of phase space
but how large is this contribution in the threshold region
$\mu \simeq \mq$?
\figxfigv

The FFS calculation (LRSN) does not treat the heavy quark as a parton.
Instead, the heavy-quark contribution is included explicitly into the
hard scattering via the flavor-creation (FC) process (\fig{xfigiii}b).
In this case, the heavy-quark contributions are {\it not} restricted in
phase space.  The splitting of the initial state gluon is
only included up to \oas2 but covers the collinear on-shell region as well
as the region where the $t$-channel heavy quark is far off-shell.
We therefore expect LRSN to provide the best results when the $t$-channel
heavy quark is highly virtual, but how much of the on-shell collinear
region does LRSN include?

We want to investigate how important the roles that the
flavor creation  (LRSN calculation) and the flavor excitation
(ACOT calculation) processes play when in the threshold region ($Q \sim \mq$),
in the intermediate transition region ($Q \, > \, \mq$), and
in the asymptotic region ($Q \gg \mq$).

\subsection{Threshold Region}

In the threshold region, we do not expect the QCD evolution of the heavy
quark PDF to make a significant contribution since the heavy-quark
evolution begins at scale $\mu \sim \mq$, and ends at scale $\mu \sim Q$ not
significantly larger than $\mq$.
More specifically, in the threshold region, the heavy quark will typically be
produced far off-shell such that the dominant region of phase space is the
virtual region.  Since the ACOT calculation picks up primarily the on-shell
region of phase space while the LRSN calculation picks up the entire phase
space, we expect the LRSN
calculation should make the better prediction in this region.

In fact, if the physical threshold for heavy-quark production is below the
threshold for introducing the heavy quark into the PDF ($\mu<\mq$), then
there will be no contribution from the heavy-quark QCD evolution since
$f_Q(\mu<\mq)=0$.  In the limit, both the \ff0  and \ff{1s}  contributions will
vanish so that  only \ff{1u} remains. It
is now evident why we defined the K-factor for the ACOT calculation by
\eq{kfacACOT}.  In this limit, the K-factor is unity, and there is no
contribution from higher orders in this kinematic regime.\footnote{
Note that by a choice of factorization
scale, we can always ensure (although sometimes artificially) that the
threshold for heavy quarks in the PDF's is always less than the
physical heavy-quark production threshold. We will discuss the
factorization scheme and the scale dependence in later sections.}
On the other hand, since the LRSN calculation computes the higher
order corrections for a virtual heavy-quark exchange, the result is a
non-trivial K-factor near threshold.

We conclude that near threshold, the dominant region of phase space is the
virtual  region.  We therefore expect the LRSN
calculation to determine more accurately the higher-order contributions in
this region.

\subsection{Asymptotic Region}

We now consider the asymptotic region where $\mu_{\rm phy} \gg \mq$.
In this region, we can essentially neglect the mass of the heavy quark in
comparison to the characteristic energy scale of the process, $\mu_{\rm phy}$.
Because the mass of the `heavy' quark is small relative to $\mu_{\rm phy}$,
the `heavy' quark is easily produced
on-shell with relatively little transverse momentum (of order $\mq$) as
compared to its longitudinal momentum (of order $\mu_{\rm phy}$) in the
$\gamma^*$-hadron center of momentum frame.

Equivalently, the QCD evolution will now be significant because there is a
large region (from $\mq$ to $\mu$) over which the evolution can build up the
heavy-quark parton distribution.
Therefore, we expect the dominant region of phase space is the collinear
region, and that the  ACOT approach should correctly resum the important
higher order contributions.

In this asymptotic limit, the collinear portion of \ff2 in LRSN
(see \fig{xfigiii}) will be contained within the ACOT result as are the other
higher-order gluon ladder graphs.  If the phase space is dominated by the
collinear region then the ACOT calculation is well suited to make accurate
predictions in the asymptotic region since the dominant terms, the infinite set
of recursive quark and gluon splittings, have been resummed.
The  singular $\alpha_s \ln(Q/\mq)$  present in the LRSN calculation
have been reorganized in ACOT (absorbed into the heavy-quark PDF) to leave
only terms of order $\alpha_s \ln(Q/\mu)$.

If it were to happen that the terms of order \oas3 and higher were a
negligible contribution to the \oas1 and \oas2 terms computed in the LRSN
calculation, then we would expect the ACOT and LRSN calculations to match in
this asymptotic region.  Conversely, the difference  between these
calculations in this region is indicative of the size of the terms of order
\oas3 and higher which have been resummed in the QCD evolution of the heavy
quark parton distributions.

While the ACOT calculation is expected to provide the more accurate results
at asymptotic $Q^2$, the question arises:  what $Q^2$ qualifies
as asymptotic?  As $Q^2$ gets large then $\alphas
\ln(Q^2/\mq^2)$ can grow to ${\cal O}(1)$ and spoil the convergence of
the perturbation series.  Investigating this problem in a cursory way, we
present the expression for $\alphas$ at the two-loop level to be
\beqn
\alphas(\mu) &=&
\frac{4 \pi}{\beta_1 \ln \left( \mu^2 / \Lambda_{QCD}^2 \right) } \
\left[
1 - \frac{\beta_2 \ln \ln \left( \mu^2 / \Lambda_{QCD}^2 \right)}
{\beta_1^2 \ln \left( \mu^2 / \Lambda_{QCD}^2 \right) }
\right] \, ,
\eeqn
where $\beta_1$  and $\beta_2$ are given by
\beqn
\beta_1 &=& \frac{11}{3} C_A - \frac{4}{3} n_f \, t_F \, ,
\\[5pt]
\beta_2 &=& \frac{34}{3} C_A^2 - \frac{20}{3} C_A n_f \, t_F
- 4 C_F n_f \, t_F \, ,
\eeqn
 $n_f$ is the number of light flavors,
 $C_A = N $, $C_F = (N^2 - 1)/2N$,
 $t_F = 1/2$ for QCD,
 and
$N$ is from the $SU(N)$ nature of QCD.
 We see from this expression for $\alphas$ that
while $\ln (Q/\mq)$ is growing with increasing $Q$, this effect will
be offset by the diminishing of $\alphas$ with the dominant
$\ln(\mu / \Lambda)$ term in
the denominator since $\mu$ must be chosen as a function of $Q$ and/or
$M_Q$.  Thus it is not clear what $Q$ we can
consider to be asymptotic, and we need to analyze the results
of this comparison to draw definite conclusions.

\subsection{Intermediate Region}

The intermediate region is the most interesting as the heavy-quark production
process is a complex interplay of all available mechanisms.
Complications arise because we have no limiting
behavior to guide us as we analyze the results of each calculation.
What we do know is that the LRSN calculation should provide
the most accurate results in the threshold region and ACOT should in the
asymptotic.  What remains to be seen is how and where the transition occurs
from one to the other.  We have little intuition about this region, and
must rely on the comparison of these two calculations to give us insight
into the physics.

One portent of future progress in heavy-flavor production, the merging of
the two calculations would be most effective in this region, as neither
the flavor excitation process nor the flavor creation process should be
the dominant mechanism.  The three-order calculation that combines
the ACOT and LRSN calculations into one should have the
virtues of both approaches, have considerably reduced mass factorization scale
dependence, and allow us to make very accurate predictions to compare with the
data from HERA.  This effort is currently underway~\cite{aort}.

\section{Comparison}

We present our results using
the CTEQ2 PDF set \cite{cteq2}, which begins the charm quark QCD evolution
$\mu_0 = m_c = 1.6$ GeV,
and the bottom quark QCD evolution at
$\mu_0 = m_b = 5.0$ GeV,
 For $\mu \leq \mu_0$,  the heavy-quark density in the hadron vanishes,
$f_Q(x,\mu \, < \, \mq) = 0$.
 For the scale $\mu$, we make the choice
\beqn
\mu^2 =
\cases{
\mq^2 & if $Q^2 < \mq^2$  \cr
\mq^2 + c Q^2 (1 - \frac{\mq^2}{Q^2})^n & if $Q^2 \geq \mq^2$ \cr}
\eeqn
with $c = 1/2$ and $n = 2$ as discussed in ACOT.

\subsection{C-Quark Production}

\subsubsection{Charm $F_2$ {\it vs.} $\{x,Q^2\}$}

We now present a series of plots showing the $F_2$'s
for various values of $Q^2$ {\it vs.} $x$.
The $F_2$ K-factors are shown to facilitate comparison between the different
curves.
The values of the $F_2$ structure functions are also shown on a log-log plot
to  indicate difference in an absolute sense, and to gauge the contribution
to the cross section which is proportional to the integral of $F_2$ over $x$.

We refer to the curves as FE for the \oas0 flavor excitation  process,
FC for the \oas1 flavor creation process, ACOT for the complete VFS
calculation and LRSN for the complete FFS calculation.\footnote{
Although we made a thorough comparison throughout the kinematic regime,
for sake of space, only characteristic plots are shown.
The $\{x,Q^2\}$ range was investigated in uniform logarithmic steps, which
explains the odd choice of $Q$ values presented.}
The FC curve is not shown as this is the denominator in the
definition of the K-factors. (Trivially, $K_{FC}=1$.)

For $Q = 1.778 $ GeV, (\fig{cxi}) the FE K-factor essentially vanishes as
compared to the other contributions because $f_c(x,\mu) \simeq 0$.
This result matches our expectation that ``heavy-quark" partons  should not
contribute at low energy scales.
The ACOT K-factor is approximately equal to one; that is, the ACOT reduces
to the FC result in the low energy limit.  This is because we have
$f_c(x,\mu)
\simeq f^g_N(x,\mu) \otimes f_g^{c \, (1)}(x,\mu)$  in the threshold region
causing
$(\ff0_{\scriptscriptstyle V} - \ff{1s}_{\scriptscriptstyle V}) \simeq 0$,
and therefore $K_{ACOT} \simeq 1$.
The rise in the FE and ACOT K-factors is due to the difference between the
one-particle and two-particle phase space factors in the threshold region
($x \to x_{max}$). The one-particle final state requires the partonic
center-of-mass energy $\sqrt{s} > M_Q$ and the two-particle final state
demands $\sqrt{s} > 2 M_Q$.  This causes the K-factors for ACOT to
diverge for $2 M_Q > \sqrt{s} > M_Q$, where
$s = Q^2 (1-x)/x$.  The rise in the LRSN K-factor is attributed to the
dominance of the \oas2 contributions over the
\oas1 result, as can be seen at the partonic level in Figures 6a and 7a of
\cite{lrsn}.
The effect on the absolute value of $F_2$  and the cross
sections is negligible as can be seen in \fig{cxi}b.
Clearly, in this low $Q$-region, the LRSN calculation is the most
appropriate because the FC photon-gluon fusion process is the dominant one, the
\oas2 corrections to that channel are included in the LRSN
calculation and provide a non-trivial K-factor.  The corrections at large $x$
are primarily a result of the Coulomb exchange of a gluon between the
final-state heavy quark and antiquark as well as initial-state-gluon
bremsstrahlung.

\figcxi
\figcxii
\figcxiii

At $Q =  5.623 $ GeV (\fig{cxii}), the FE $F_2$ result has increased
significantly and approaches the other curves.
This very fast evolution of the FE result makes it unreliable for
predicting  heavy-quark production.\footnote{
Note that for larger values of $Q^2$, the FC process appears to match roughly
the LRSN, and to match very well with the ACOT calculation.
This agreement is fortuitous and depends on a judicious choice
for the factorization scale $\mu$. As we shall see shortly, the FC result
is very sensitive to $\mu$.}
The ACOT K-factor now deviates from unity as the imperfect cancellation
between $\ff0_{\scriptscriptstyle V}$ and $\ff{1s}_{\scriptscriptstyle V}$
signals the existence of non-trivial contributions from the heavy-quark PDF
evolution.
The fast evolution of the heavy quark in the threshold region (due to
abundant gluons) generates important contributions for relatively low
values of $Q$.  However the subtraction prescription ensures
the result is reliable (in contrast to the FE process), as we shall confirm
when we examine the $\mu$-dependence.
The LRSN K-factor decreases at small $x$, and flattens slightly.
This decrease at small $x$ is due in large part to the cancellation between
the mass factorization logarithmic terms and the mass factorization scale
independent terms at large parton-photon center-of-mass energies $s$.
 For additional discussions, see pp.192-196 of reference~\cite{lrsn}.

At $Q =  17.78 $ GeV (\fig{cxiii}), the FE, ACOT, and LRSN results have
similar shapes.   The K-factors are all  monotonically increasing vs. $x$,
with a large rise in the threshold ($x\to x_{max}$) region.
In the range above $Q = 17.78 \,GeV$, the general characteristics are
similar to this \fig{cxiii}.

\subsubsection{Charm $F_2$ {\it vs.} $\mu$}

We now present a series of plots of the $\mu$-dependence of $F_2$ for
particular $x$ and $Q^2$ values accessible to HERA.

For $x = 0.001$ and $Q = 3.162 $ GeV (\fig{cmui}), the FE result has a large
$\mu$-dependence. The FE process is driven by the heavy-quark PDF and vanishes
at small $\mu$ because  $f_c(x,\mu) = 0$ for $\mu \leq
m_c$. As $\mu$ increases above $m_c$, $f_c(x,\mu)$ increases quickly due to
the abundance of gluons and the dearth of heavy quarks near threshold.
Clearly, the FE result is extremely sensitive to  the particular choice
of scale, as well as the choice of factorization scheme which determines
where to start the heavy-quark evolution.
As anticipated, the apparent agreement between the FE result and the
other results as seen in the previous subsection is merely an accident due
to a prudent choice of scale, and cannot provide stable results.
The LRSN result is quite flat, and the ACOT calculation exhibits a rather
marked dependence on the scale choice, although substantially less than the FE
result.

\figcmui
\figcmuii
\figcmuiii
\figcmuiv

For $x = 0.001$ and $Q = 10 $ GeV (\fig{cmuii}),
the ACOT and the LRSN calculations exhibit a comparable and small dependence
on the scale choice. In this region,  neither is a clear improvement over
the FC result.

For $x = 0.01$ and $Q = 10 $ GeV (\fig{cmuiii}),
both the LRSN and ACOT calculations are essentially flat,  and both
appear to be an improvement over the FC result.

For $x = 0.1$ and $Q = 31.62 $ GeV (\fig{cmuiv}),
the ACOT calculation exhibits slightly less dependence
than the LRSN result, and both are considerably less scale-dependent than
the FC and FE  results.

The general trend revealed here confirms the reduced $\mu$-dependence of the
ACOT and LRSN results over their one-order counterparts.  Furthermore, we also
have confirmation that LRSN is the appropriate calculation near threshold
and ACOT in the asymptotic region, as shown when $Q \gg \mq$.

\subsection{B-Quark Production}

\subsubsection{Bottom $F_2$ {\it vs.} $\{x,Q^2\}$}

We have also compared the ACOT and the LRSN results for b-quark production.
We find very similar features in \fig{bxi} with $Q = 5.623 $ GeV
for $b$ production as we did in \fig{cxi} with $Q = 1.778 $ GeV for $c$.
This is as expected since the physics is set by the relevant mass ratio in the
process: $\mq/\mu$.  The LRSN calculation has a non-trivial K-factor,
the FE result essentially vanishes, and the ACOT K-factor is essentially
unity.

\figbxi
\figbxii
\figbxiii

For $Q = 10 $ GeV (\fig{bxii}), the FE results are roughly comparable
to the other curves as the heavy-quark evolution turns on.  The ACOT K-factor
deviates from the FC result as the difference between
$\ff0_{\scriptscriptstyle V}$ and $\ff{1s}_{\scriptscriptstyle V}$ again
signals the existence of non-trivial contributions from the heavy-quark
PDF evolution.  The LRSN  K-factor decreases at small-$x$, and flattens
slightly again for the reasons discussed concerning \fig{cxii}.

In \fig{bxiii}, we display the results for $Q = 31.62$ GeV.
The ACOT  exhibits a significant difference from the FC process as the
heavy-quark evolution continues.  As before,  the FE, ACOT, and LRSN
results have similar shapes, and each K-factor is
monotonically increasing as $x$ increases.  In the range above $Q = 31.62 $
GeV, the general characteristics are similar to this figure.

\subsubsection{Bottom $F_2$ {\it vs.} $\mu$}

We now examine the $\mu$-dependence of $b$ production at various
points in $x$ and $Q^2$ space relevant to the region accessible at HERA.

For $x = 0.001$ and $Q = 10 $ GeV (\fig{bmui}), the FE result again
has a large $\mu$-dependence because this result is closely tied to
$f_b(x,\mu)$.
The ACOT result is a clear improvement over the FE result, but the FC result
exhibits still less dependence and is comparable with the LRSN calculation.

\figbmui
\figbmuii
\figbmuiii

For $x = 0.01$ and $Q = 10 $ GeV (\fig{bmuii}),
the FE result still has a large $\mu$-dependence, and
both the ACOT and the LRSN results improve upon the FC result.

For $x = 0.1$ and $Q = 31.62 $ GeV (\fig{bmuiii}),
the LRSN and the ACOT calculations are comparable and both show
substantially less $\mu$-dependence than the FE or FC
results.

We can make some general observations regarding the $\mu$-dependence for both
charm and bottom production.
For all values of $Q$, the FE process is increasing with  $\mu$ due to
the increasing heavy-quark PDF.  In contrast, the \oas1 FC process (driven by
gluons) is decreasing with $\mu$ largely due to the decrease in $\alphas(\mu)$.
The two-order calculations (ACOT and LRSN) that have compensating
contributions to cancel out some of the $\mu$-dependence.
Specifically, ACOT combines pieces of the FE and FC processes (together with a
subtraction term) to yield a result that has substantially less
$\mu$-dependence than either result in the large $Q$ region.  LRSN effectively
has the \oas2 FE contribution as the collinear heavy-quark part of phase
space is included, negating some of the $\mu$-dependence of the FC channel.

We note that in general, the ACOT result exhibits minimal $\mu$-dependence
at larger values of $Q$ and $x$.  At large $Q$, we are closer to the
asymptotic region where the VFS approach is expected to be superior.
The increased $\mu$-dependence at low $x$ arises mainly from the quark-initated
contributions. At a given order, we expect the quark-initiated contributions
to be less than the gluon-initiated ones because
$f_q(x,\mu) \ll f_g(x,\mu)$~\cite{acot}.
However, because the second-order Altarelli-Parisi splitting kernels
$P_{q\to q}^{(2)}(x)$ and $P_{g\to q}^{(2)}(x)$ contain singular $1/x$ terms
whereas the first-order $P_{q\to q}^{(1)}(x)$ and $P_{g\to q}^{(1)}(x)$ do not,
the evolution of the quark distribution at small-$x$ will be completely
dominated by the second-order kernel rather than the first-order kernel
\cite{wkt}.  Consequently, contributions from higher order quark-initiated
processes should cancel the above $\mu$-dependence. This work is in progress
\cite{aort}.

In examining the
$\mu$-dependence plots, the reader may have noticed for some $x$ and $Q^2$
values chosen, particularly at small $x$, the results of ACOT and LRSN
never crossed.  The conventional wisdom says the $\mu$-dependence
represents the uncertainty from uncalculated higher-order contributions.
What are the implications when the $\mu$-dependence curves do not
cross when a significant region along the $\mu$-axis has been traversed?
 At lower values of $Q^2$, the difference between ACOT and LRSN is expected
due to the limited evolution of the heavy quark PDF.
 At asymptotic values of $Q^2$, the difference between ACOT and LRSN is within
the
range suggested by the $\mu$ dependence.
 However, in the intermediate range,  the range of the
$\mu$-dependence may underestimate the full theoretical uncertainty.

\subsubsection{Total Charm and Bottom Cross Section}

To estimate the theoretical uncertainty for the total cross section, we
have  computed
\beqn
F_2(Q^2) = \int_{10^{-4}}^{x_{max}} dx F_2(x,Q^2)
\label{eq:G}
\eeqn
which is proportional to the dominant contribution for the heavy-quark
production cross section and show the ratio of $F_2^{ACOT}/F_2^{FC}$ and
$F_2^{LRSN}/F_2^{FC}$ for $c$ production in \fig{cint}.  For low $Q$, the
differences are on the order of 30\%.  As $Q$ increases, however, the
differences decrease to $\simeq 12\sim15\%$ as $Q \simeq 100 $ GeV.  Similar
behavior occurs for $b$ production, as seen in \fig{bint}.
 Some of this difference may possibly be attributed to choice of scale or
PDF set.  However variations in scale or PDF set should not be solely
responsible for this difference.  We may have to await the data from HERA
to resolve this discrepancy.

\figcint
\figbint

\section{Conclusions}

We have outlined the strengths and weaknesses of both the  VFS (ACOT) and
the FFS (LRSN) calculation. We summarize the highlights below.

\begin{itemize}

\item
In the threshold region, the FFS (LRSN) calculation yields the most stable and
reliable results due to the domination of flavor creation.

\item
In the asymptotic region, the VFS (ACOT)  calculation provides the best
results because of the dominance of the collinear heavy-quark contribution.

\item
In the kinematic range spanned by HERA, the
$\alphas \log[Q^2/\mq^2]$ terms are not a significant problem for
the FFS (LRSN) calculation.

\item
In the VFS (ACOT) calculations, the heavy-quark PDF's can yield
significant contributions at relatively small scales,
({\it i.e.} $\mu/\mq \sim 3$).

\item
While the flavor excitation (FE) process can closely match the two-order
results with a judicious choice of the scale $\mu$, the large scale dependence
makes this unreliable in computing structure functions.

\item
Likewise, while the flavor creation (FC) process is a good starting point
in the threshold region.  However, the LRSN calculation indicates that the
corrections to this na{\'i}ve estimate can be as large as 100\%.

\end{itemize}

The range of $x$ and $Q^2$ we have presented reflect the region accessible at
HERA.
 We note that the difference between the LRSN and ACOT calculations above
threshold is suggestive of higher order contributions yet to be included.
 As such, the results of this comparison indicate that a combining of the LRSN
and ACOT calculations  in a consistent fashion (with  the additional mass
factorizations required) should allow us to make predictions based upon a
three-order result that combines the best attributes of both calculations.
 The result should be a calculation that will provide an important test of
perturbative QCD  when compared with the results from HERA.

\section*{Acknowledgements}

The authors would like to thank John Collins, Jack Smith, Davison Soper and
Wu-Ki Tung for useful discussions.   This work is partially supported by
the  U.S. Department of Energy Contract No. DE-FG05-92ER-40722,
by the Texas National Research Laboratory Commission,  and
by the Lightner-Sams Foundation. F.O. is supported in part by an SSC
Fellowship.


\end{document}